\documentstyle[sprocl,epsfig]{article}

\def\Journal#1#2#3#4{{#1} {\bf #2}, #3 (#4)}
\def\Ap{\em ApJ}

\def\A\&A{\em A\&A}
\def\AR{\em ARA\&A}
\def\Pa{\em PASP}
\def\ApS{\em ApJS.}

\def\be{\begin{equation}}
\def\ee{\end{equation}}
\def\bea{\begin{eqnarray}}
\def\eea{\end{eqnarray}}

\begin{document}

\title{OPTICAL FeII EMISSION IN AGN}

\author{C. BONGARDO, R. ZAMANOV, P.MARZIANI, M. CALVANI}

\address{INAF, Osservatorio Astronomico di Padova, Italy}

\author{J.W. SULENTIC}

\address{Department of Physics and Astronomy, University of Alabama,
     Tuscaloosa, USA}

\maketitle

\abstracts{We investigated the optical FeII emission in a sample of about
215 low-redshift AGN (quasars and luminous Seyfert 1 galaxies). We find that
a scaled and broadened FeII template based on the I Zw 1 spectrum can
satisfactorily model the FeII emission in almost all sources in our sample.
We confirm that FWHM H$\beta$ and FeII$\lambda$4570 are strongly
correlated.  The correlation is different for sources with FWHM(H$\beta$)
greater than or less than $\sim$4000 km s$^{-1}$.  Sources with FWHM
H$\beta$$\leq$ 4000 km s$^{-1}$ (Population A) show no difference between
FWHM H$\beta$ and FeII while sources with FWHM(H$\beta$) $\geq$ 4000 km
s$^{-1}$ (Population B) show FWHM FeII that is systematically smaller than
FWHM H$\beta$.  This may be telling us that FeII emission in Pop. B sources
comes from only the outermost part of the H$\beta$\ emitting region where
the degree of ionization is lowest.}

\section{Introduction}

FeII emission has not been fully exploited as an emission line diagnostic,
because the complexity of the Fe$^+$ atom hampers reliable modeling. It was
long ago realized that FeII is emitted in the BLR and that strong FeII
emission requires a high density (n$_e$ $\geq$ 10$^9$ cm$^{-3}$, high
column density and low temperature (T$_e$ $\approx$ 5000 K; [1]) emitting
region. These conditions are thought occur in a partially ionized region
within the BLR created by strong X-ray emission ([2], [3] and [4]). The
principal observational evidence for FeII emission from the BLR involves
the profile (i.e. FWHM) similarity between optical FeII emission lines and
the broad component of H$\beta$ ([5] and [6]). We show that the correlation
between FWHM(FeII$\lambda$4570) and FWHM(H$\beta$) is highly significant
for a large sample of  low redshift AGN; however, it takes a different form
for sources with FWHM(H$\beta$) greater or less than $\sim$4000 km s$^{-1}$.

\section{The Template}

We derived FeII emission properties for each source using a scaled and
broadened template extracted from a high S/N spectrum of I Zw 1. This
method has already been successfully applied ([7], [8]). The template
allows us to satisfactorily model and extract the FeII emission in about
98\%\ of our sample. In other words, we found very few cases where the FeII
line and multiplet ratios deviated from the I Zw 1 template.  Fig. 1 shows
examples of FeII$_{opt}$ emission extraction in sources with very different
broad line width.

\begin{figure}
\vskip -0.7 true cm \centerline{\epsfig{figure=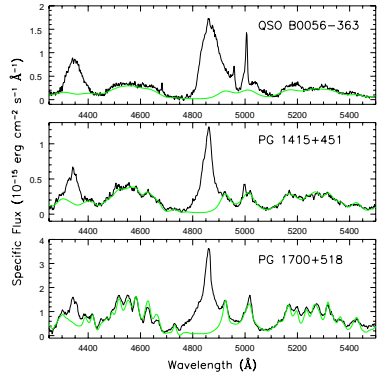, width=5cm,
height=5cm}} \vskip -0.3 true cm \caption{Spectra of the H$\beta$ region
with a superimposed model of FeII$_{opt}$. (thick green line).}
\end{figure}

\section{FWHM(FeII$\lambda$4570) vs. FWHM(H$\beta$)}

Measurement of FWHM for optical FeII is not trivial. It is necessary to
perform the following steps in order to parameterize optical FeII emission.
(1) Define the local continuum using regions around 4200, 4700 and 5000 \AA
that are relatively free of FeII contamination. (2) Subtract an adopted
optical FeII template that has been scaled and broadened to give minimum
residuals in the FeII$\lambda$4570\AA. (3) Determine the template
broadening factor that is proportional to FWHM(FeII$\lambda$4570) because
the line FWHM is much larger than the instrumental profile. FeII equivalent
width and FWHM measurement accuracy depends on S/N ratio and intrinsic
width of the emission. Given a fixed S/N, there is a minimum W(FeII) for
which the blends become undetectable (i.e., features are lost in noise or
create a pseudo-continuum). FeII equivalent width measurements were
obtained for 112 objects with an upper limit given for other sources. FWHM
was derived for most, but not all, of the 112 sources. We usually estimated
the FWHM uncertainty by evaluating a matrix where residuals were shown as a
function of FeII intensity and broadening factor. FWHM boundaries were set
when residuals were significantly worse than for the best fit. It was not
always possible to constrain the width in sources where EW FeII was small.

In order to accurately measure  FWHM(H$\beta_{BC}$)  we also subtracted
[OIII]$\lambda\lambda$ 4959, 5007\AA emission lines by interpolating
between the blue and red edges. The broad and/or narrow components of Broad
and/or narrow components of HeII$\lambda$4686\ were subtracted, when
present, using a Gaussian fit. Fig. 2 shows a correlation plot for
FWHM(FeII$\lambda$4570) vs. FWHM(H$\beta$).  Outlying data points most
likely reflect poor S/N, low resolution and/or FeII  weakness rather than
real differences. Only one outlier source is convincing: IRAS 07598+6508
(see \S \ref{iras}). The correlation takes a different form if
FWHM(H$\beta$) $\leq$ 4000 km s$^{-1}$ (Population A,  following [9]) or
FWHM(H$\beta$) $\geq$ 4000 km s$^{-1}$ (Population B):
\begin{itemize}
\item Pop. A. There is a reasonably strong linear correlation (r$_{Pearson}=0.69,\ N=69,\
P=8.5\times10^{-9}$) supporting the assumption that FWHM(H$\beta$)=
FWHM(FeII$\lambda$4570) and implying that both arise in  the same emitting
region.
\item Pop. B. There is also a clear correlation
(R$_{Pearson}=0.88,\ N=43,\ P=7.3\times10^{-9}$), however
FWHM(H$\beta_{BC}$) appears to be systematically broader than
FWHM(FeII$\lambda$4570).
\end{itemize}

\begin{figure}
\begin{tabular}{cc}
\epsfig{figure=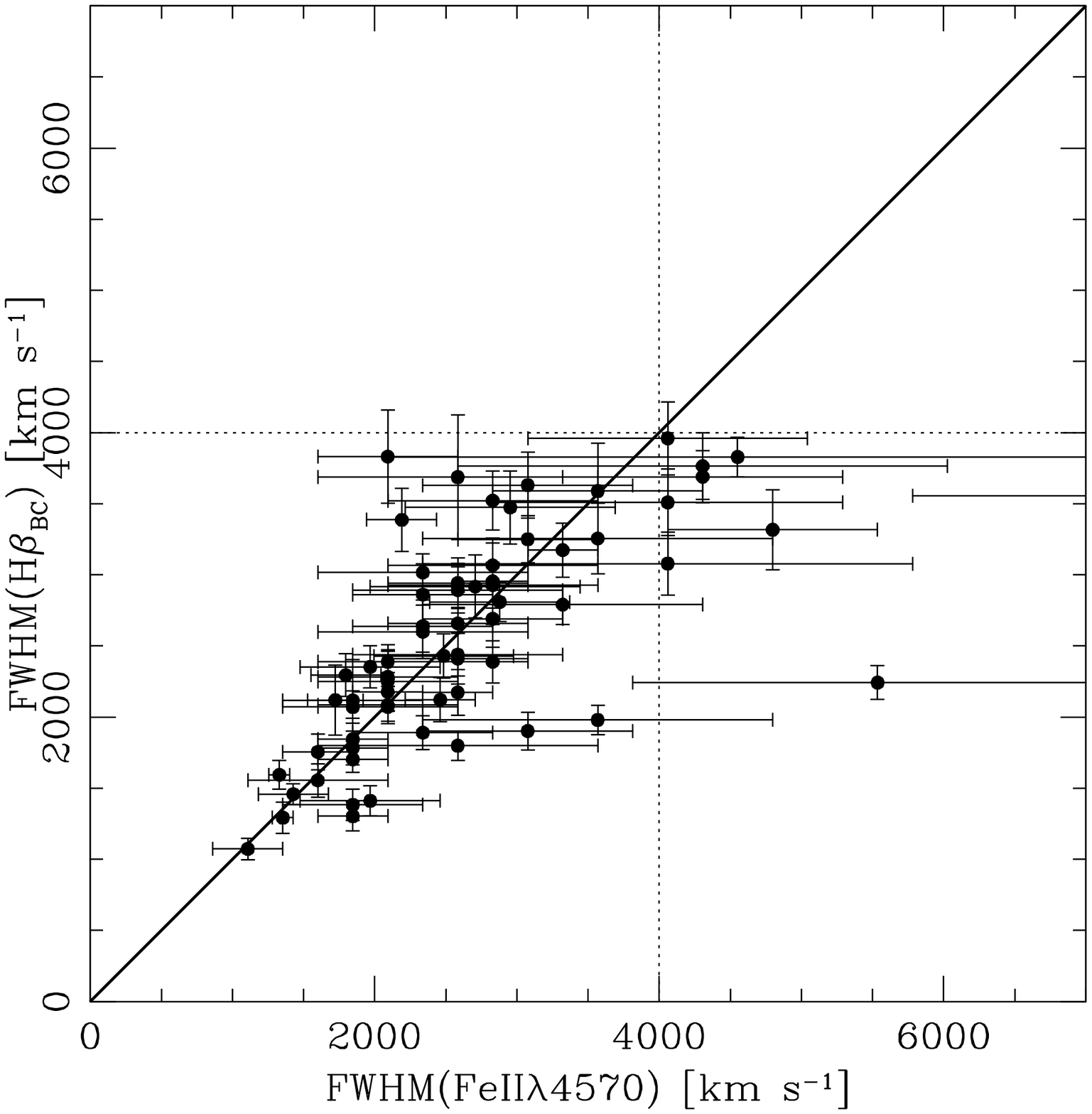, width=5cm, height=5cm}
&\epsfig{figure=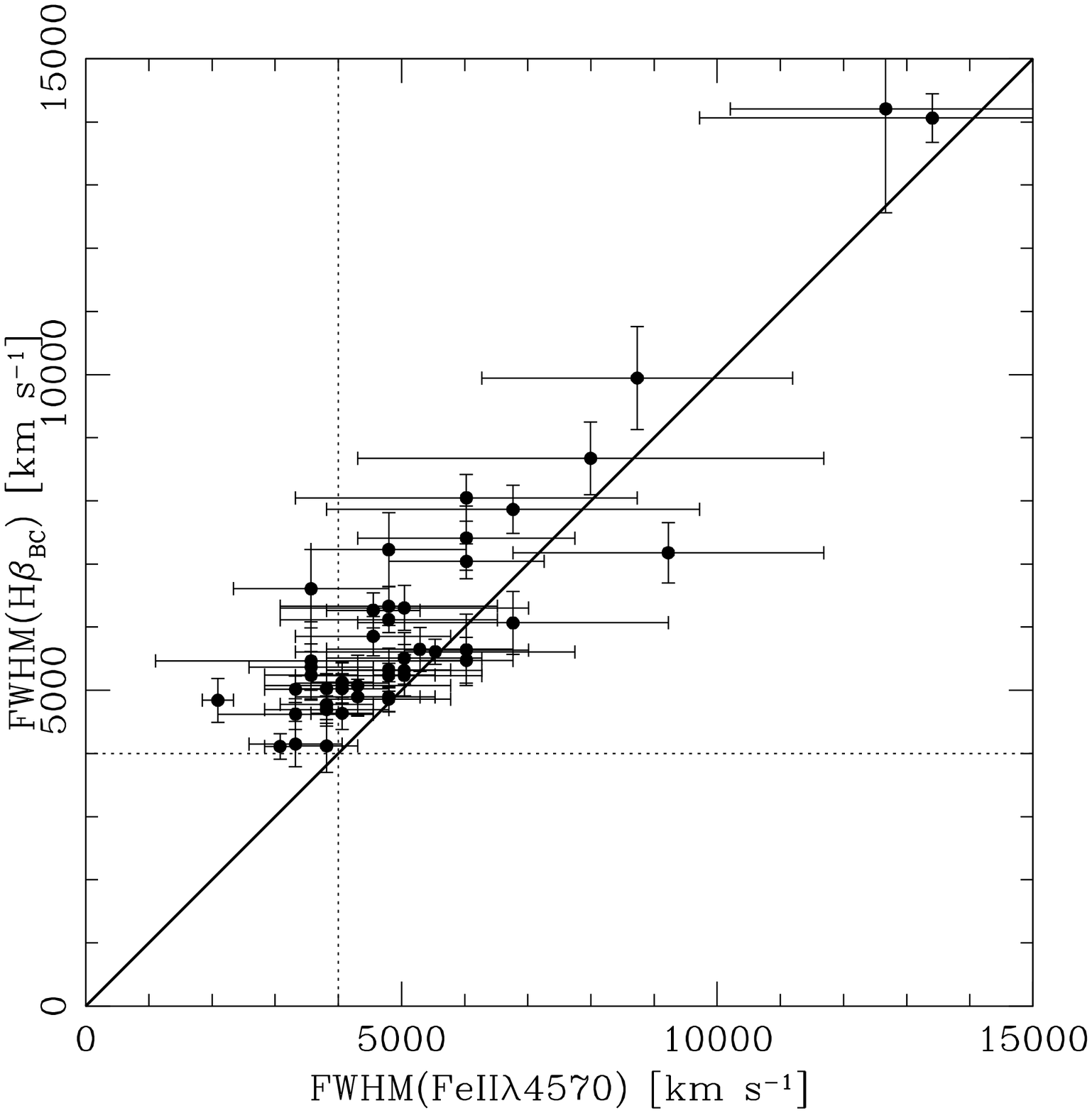, width=5cm, height=5cm}
\end{tabular}
\caption{Correlation plots FWHM(H$\beta_{BC}$) vs. FWHM(FeII$\lambda$4570)
for Population A (left) and Population B (right) sources.}
\end{figure}

\subsection{IRAS 07598+6508: a special case? \label{iras}}

This intriguing FIR excess AGN source shows an unusual location in our
optical Eigenvector 1 diagram ([10]). FWHM(H$\beta_{BC}$)=5000$\pm$400 km
s$^{-1}$ and FWHM(FeII$\lambda$4570)=2000$\pm$1300 km s$^{-1}$ in this
source.  The high S/N spectra and strength of the FeII$_{opt}$ emission,
along with the large EW(H$\beta_{BC}$), make this result especially
striking. The strong blueward asymmetry of the BC in H$\beta$ (see fig. 3)
suggests that additional broadening is present, due probably to Balmer
emission from a highly blueshifted CIV$\lambda$1549 \AA\  emitting  wind. A
narrower and unshifted component can be associated with low ionization
emission typical of the majority of Type 1 AGN.

\begin{figure}
\vskip -0.7 true cm \centerline{\epsfig{figure=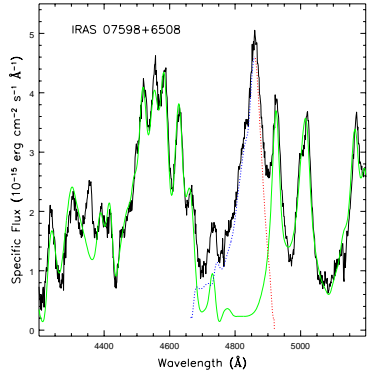, width=5cm,
height=5cm}} \vskip -0.3 true cm \caption{H$\beta$ spectral region of IRAS
07598+6508. The dotted blue/red line represents the H$\beta_{BC}$.}
\end{figure}

\section{Conclusion}
The most straightforward implications of our results may be that:  1) the
FeII$_{opt}$\ emission mechanism is probably the same in almost all AGN and
2) FeII$_{opt}$\ emission is primarily emitted in a region of the BLR where
H$\beta$\ produced. FWHM FeII$_{opt}$ in Population B objects appears to be
systematically narrower than FWHM H$\beta$. This suggests that FeII
emission in Pop. B sources may come from only the outermost part of the
H$\beta$\ emitting region where the ionization degree is lower. This is
again evidence supporting the population A-B distinction among AGN

\section*{Acknowledgements}
We acknowledge support from the Italian MURST through Cofin 00--02--004.

\end{document}